 \documentclass[aip,jcp,reprint,showpacs,superscriptaddress]{revtex4-2}
 \usepackage{amsmath,amssymb,mathrsfs,amsfonts,dsfont}
 \usepackage{bm}
 \usepackage{color}
 \usepackage{appendix}
 \usepackage{graphicx}
 \usepackage{footmisc}
 \usepackage{bm}
 \usepackage{color}
 \usepackage{CJKutf8}
 \usepackage{orcidlink}
 \usepackage{hyperref}
 \hypersetup{colorlinks=true,
 	linkcolor=red,
 	citecolor=blue,
 	urlcolor=blue}

\begin{document}
  \title{Photon counting statistics in the presence of spectral diffusion induced by nonequilibrium environmental fluctuations}
   \author{Xiangji Cai~\orcidlink{0000-0001-6655-5736}}
  \email{xiangjicai@foxmail.com}
  \affiliation{School of Science, Shandong Jianzhu University, Jinan, 250101, China}

  \author{Yonggang Peng~\orcidlink{0000-0003-0931-3057}}
  \email[Authors to whom correspondence should be addressed: ]{ygpeng@sdu.edu.cn; and yzheng@sdu.edu.cn}
  \affiliation{School of Physics, Shandong University, Jinan, 250100, China}
  \author{Yujun Zheng~\orcidlink{0000-0001-9974-2917}}
  \email[Authors to whom correspondence should be addressed: ]{ygpeng@sdu.edu.cn; and yzheng@sdu.edu.cn}
  \affiliation{School of Physics, Shandong University, Jinan, 250100, China}

\begin{abstract}
We theoretically investigate the statistical properties of photon emission of a driven two-level single-molecule system undergoing spectral diffusion induced by nonequilibrium environmental fluctuations. Within the framework of the generating function method and the stochastic Liouville equation, we analyze the influence of the nonequilibrium characteristics of environmental fluctuations respectively governed by nonstationary Ornstein–Uhlenbeck noise  and random telegraph noise on the photon counting statistics of the driven single-molecule system.
In the slow modulation limit of spectral diffusion, the intensity and statistical fluctuations of photon emission depend on the environmental nonequilibrium characteristics at short time scales, whereas they become independent of the nonequilibrium characteristics of environmental fluctuations in the steady state.
In the fast modulation limit of spectral diffusion, neither the line shape nor the Mandel’s parameter depends on the environmental nonequilibrium characteristics owing to the rapid relaxation of environmental fluctuations.
These findings not only shed light on the role of nonequilibrium environmental fluctuations in shaping the photon emission properties of single-molecule systems but also provide a basis for distinguishing between equilibrium and nonequilibrium characteristics of environmental fluctuations in experimental measurements.
\end{abstract}


\maketitle

\section{Introduction}
\label{sec:Intr}

With the continuous advancement of our ability to produce new materials and to observe and regulate quantum systems across varying time, length, and energy scales, non-equilibrium kinetics and quantum coherent dynamics increasingly play an indispensable role in uncovering novel phenomena in physical, chemical, and biological processes.~\cite{RevModPhys.59.1,JChemPhys.104.4189,JChemPhys.106.5239,Nature446.782,JChemPhys.130.084105,Nature463.644,ProcNatlAcadSci.107.12766,JChemPhys.145.204118,RevModPhys.88.021002,RevModPhys.89.015001,SciAdv.6.eaaz4888,ChemRev.124.11641,JChemPhys.160.054101}
In many of these scenarios, single-molecule spectroscopy (SMS) has emerged as a transformative technique in laboratories worldwide. By eliminating the ensemble averaging inherent to conventional methods, it enables the investigation of physical, chemical, and biological systems with unprecedented precision.~\cite{JChemPhys.117.10938,RevModPhys.87.1153,RevModPhys.87.1169,RevModPhys.87.1183}
Over the past decades, continuous innovations in experimental methodologies have propelled the advancement of SMS, with growing anticipation for its expansion into fields such as nuclear magnetic resonance and other interdisciplinary domains. Its profound impact on elucidating molecular behaviors at the most fundamental level continues to escalate.
The SMS technique interrogates individual entities, including molecules, ions, nitrogen-vacancy centers, and quantum dots, in condensed-phase or biological environments, uncovering novel insights into light-matter interactions, single-particle dynamics, and nanoenvironmental coupling effects.~\cite{Cohen-Tannoudjibook,JChemPhys.114.5137,JChemPhys.117.10996,JChemPhys.117.11024,AnnuRevPhysChem.55.457,JChemPhys.121.6361,*JChemPhys.121.6373,Barkai_editedbook,RepProgPhys.74.106401,AnnuRevPhysChem.72.253,RevModPhys.96.025001}

Spectral diffusion stands as a cornerstone concept in contemporary SMS, underpinning the physical mechanisms that govern photon statistics in single-molecule systems. 
This phenomenon arises from stochastic fluctuations in the environment surrounding quantum emitters, including the reorientation of solvent molecules, the rearrangement of surface ligands, and phonon-induced energy shifts in solid matrices. 
As a critical dynamical process, spectral diffusion is also driven by environmental fluctuations across diverse physical scenarios, such as elastic collisions between atoms in gaseous media, or between phonons and atoms in solid-state systems.~\cite{Meystrebook}
These environmental fluctuations modulate the absorption frequency of single-molecule systems stochastically over time.
This modulation manifests as characteristic variations in average waiting time of photon emission, statistical fluctuations in the line shape, and modulations in intensity trajectories, which are experimentally observable and constitute unique ``spectral fingerprints" for elucidating the underlying nonequilibrium dynamics of the quantum emitter in the presence of environment coupling.
The phenomenon of spectral diffusion has been extensively observed in many single-molecule systems through a variety of experimental techniques.~\cite{Nature349.225,JChemPhys.95.7150,Nature385.143,Science282.1877,PhysRevLett.70.3584,AnnuRevPhysChem.49.441,PhysRevLett.82.2195,PhysRevLett.85.3301,JChemPhys.114.6843}
This widespread occurrence underscores its fundamental role as a signature of the interaction between individual molecules and their local environments, making it a crucial research focus in various fields.
In the past few decades, the photon emissions of driven single-molecule systems under the influence of spectral diffusion induced by environmental fluctuations in equilibrium have been extensively studied in numerous prior theoretical investigations.~\cite{PhysRevLett.90.238305,JChemPhys.119.11814,JChemPhys.121.3238,JChemPhys.121.7914,PhysRevLett.93.068302,JChemPhys.122.184703,JChemPhys.126.104303,ApplPhysLett.92.092120,PhysRevA78.015402,JChemPhys.130.244502,PhysRevA80.043831,PhysRevA83.013810,PhysRevA88.013425,PhysRevApplied22.014035}

As a matter of fact, there are numerous important scenarios where the nonequilibrium characteristics of the environmental fluctuations exert a crucial influence on the dynamical evolution of quantum systems. 
For instance, certain transient and ultrafast dynamical processes in physical, chemical and biological systems, can take place on sufficiently short time scales, whereas the initial nonstationary states of the environment induced by its interaction with the quantum system may not have the opportunity to relax to equilibrium rapidly.
In recent decades, the dynamics of open quantum systems induced by nonequilibrium environmental fluctuations have attracted widespread attention.~\cite{JChemPhys.133.241101,JChemPhys.139.024109,PhysRevA87.032338,PhysRevA94.042110,JChemPhys.149.094107,Entropy25.634,NewJPhys.22.033039,PhysRevA104.042417,PhysRevA110.012429,PhysRevA108.062208,PRXQuantum3.020321,PhysRevRes.4.013230}
Particularly in the field of SMS, the non-equilibrium characteristics of environmental fluctuations can directly give rise to a dynamically evolving spectral line shape that is distinctly different from that induced by equilibrium environmental fluctuations.

In this paper, we theoretically investigate the photon counting statistics of a two-level single-molecule system driven by an external laser field under the influence of spectral diffusion, with a particular focus on the nonequilibrium environmental fluctuations.
We analyze in detail how the nonequilibrium characteristics of environmental fluctuations affect the statistical properties of the photon emission of the single-molecule system in the presence of spectral diffusion governed by nonstationary Ornstein-Uhlenbeck noise (OUN) and random telegraph noise (RTN), respectively. 
We further compare these results with those obtained under equilibrium environmental fluctuations with stationary statistical properties.
The results show that at short timescales in the slow modulation limit, the nonequilibrium characteristics of environmental fluctuations are crucial for photon counting statistics. 
In contrast, when the modulation is fast or the observation times are long, the nonequilibrium characteristics do not measurably affect photon emission statistics, as the environmental fluctuations have already rapidly relaxed to equilibrium.

This paper is organized as follows. In Sec.~\ref{sec:TherFram1}, we briefly introduce the theoretical framework of the generating function to address the photon counting in a two-level single-molecule system driven by an external laser field, as well as the statistical characteristics of its photon emission.
In Sec.~\ref{sec:TherFram2}, we study the photon counting statistics of the driven single-molecule system under the influence of spectral diffusion induced by nonequilibrium environmental fluctuations within the framework of the stochastic Liouville equation.
In Sec.~\ref{sec:ResuDisc}, we illustrate the numerical results on how the nonequilibrium characteristics of environmental fluctuations affect the statistical properties of the photon emission 
of the driven single-molecule system.
In Sec.~\ref{sec:Conc}, we draw the conclusions from the present study.

\section{Photon counting statistics of a driven quantum system}
\label{sec:TherFram1}
We consider a two-level single molecule system with the intrinsic frequency difference $\omega_{0}$ between the excited state $|e\rangle$ and ground state $|g\rangle$, driven by an external laser field of frequency $\omega_{L}$. 
Within the rotating wave approximation (RWA), i.e.,  $|\omega_{0}-\omega_{L}|\ll\omega_{0}+\omega_{L}$~\cite{Scullybook} and making a rotating frame with the frequency of the laser field, the total Hamiltonian of the single-molecule system interacting with the laser field can be written as
	\begin{equation}
		\label{eq:Hami}
		\mathcal{H}=\frac{\hbar}{2}(\Delta_{0}\sigma_{z}+\Omega_{0}\sigma_{x}),
	\end{equation}
where $\sigma_x$ and $\sigma_z$ are the Pauli matrices, $\Delta_{0}=\omega_{0}-\omega_{L}$ is the detuning between the system and the laser and $\Omega_{0}$ denotes the Rabi frequency of the driven transition.

The dynamical evolution of the driven two-level single molecule, when interacting with a quantum radiation field initially in its vacuum state, is phenomenologically governed by the quantum master equation~(see Appendix~\ref{sec:appAmasequ})
\begin{equation}
\label{eq:evol}
  \frac{d}{dt}\rho(t)=-\frac{i}{\hbar}[\mathcal{H},\rho(t)]+\Gamma\bigg[\sigma_{-}\rho(t)\sigma_{+}-\frac{1}{2}\{\sigma_{+}\sigma_{-},\rho(t)\}\bigg],
\end{equation}
where $\Gamma$ denotes the spontaneous emission rate due to the vacuum fluctuations of the quantum radiation field, and $\sigma_{\pm}$ are the raising and lowering operators of the system.
It is worth noting that the RWA is valid under near-resonance and weak-driving conditions, whereas its accuracy degrades in the far-off-resonant case, and it breaks down under strong-field driving.~\cite{Cohen-Tannoudjibook,RevModPhys.91.025005}
Recently, investigations of driven dynamics beyond the RWA have attracted growing attention, with particular emphasis on the significant role of counter-rotating coupling on the time evolution of driven quantum systems.~\cite{PhysRevA80.033846,PhysRevA85.053830,PhysRevA86.023831,PhysRevA87.053837,PhysRevA88.053821,PhysRevLett.123.133603}

Owing to the combined effects of the external field driving and spontaneous emission, the two-level single-molecule system continuously emits photons.
The photon emission statistics of the driven single-molecule system are closely associated with the $n$th partial density matrix $\rho^{(n)}(t)$ which depends on the number of quantum jumps of the system from the exited state $|e\rangle$ to the ground state $|g\rangle$.
By means of the quantum jump method, we can obtain the time evolution of the $n$th partial density matrix as~(see Appendix~\ref{sec:appApardenmat})
\begin{equation}
	\label{eq:evopardenmat}
	\frac{d}{dt}\rho^{(n)}(t)=\mathcal{C}\rho^{(n)}(t)+\mathcal{J}\rho^{(n-1)}(t),
\end{equation}
where the operator $\mathcal{C}(\cdot)=-(i/\hbar)[\mathcal{H}_{\mathrm{eff}}(\cdot)-(\cdot)\mathcal{H}_{\mathrm{eff}}^{\dag}]$ quantifies the deterministic evolution of the single molecule governed by the non-Hermitian effective Hamiltonian $\mathcal{H}_{\mathrm{eff}}=\mathcal{H}-(i/2)\hbar\Gamma\sigma_{+}\sigma_{-}$, the operator $\mathcal{J}(\cdot)=\Gamma\sigma_{-}(\cdot)\sigma_{+}$ describes the quantum jump  from the exited state $|e\rangle$ to the ground state $|g\rangle$, and the index $n$ ranges over all integers with $\rho^{(n)}(t)=0$ for $n<0$.

Consequently, the probability that $n$ photons have been spontaneously emitted by time $t$ is given by the trace of the $n$th partial density matrix~\cite{PhysRevA23.1243}
\begin{equation}
	\label{eq:emipro}
	p_{n}(t)=\mathrm{Tr}\big[\rho^{(n)}(t)\big]=\rho_{ee}^{(n)}(t)+\rho_{gg}^{(n)}(t),
\end{equation}
and certain statistics of the photon emission of the driven two-level single-molecule system, for instance, the mean number of emitted photons, can be expressed as
\begin{equation}
	\label{eq:emipro}
	\langle N\rangle (t)=\sum_{n}np_{n}(t)=\sum_{n}n\big[\rho_{ee}^{(n)}(t)+\rho_{gg}^{(n)}(t)\big].
\end{equation}

To further investigate the statistical properties of the photon emission of the two-level single-molecule system driven by an external laser field, we define the relevant generating function within the theoretical framework originally established by Zheng and Brown~\cite{PhysRevLett.90.238305,JChemPhys.119.11814}
\begin{equation}
	\label{eq:reddengenfun}
	\rho(s,t)=\sum_{n=0}^{\infty}s^{n}\rho^{(n)}(t),
\end{equation}
where $s$ is a counting variable related to the $n$ photons emission events in the time interval $(0,t)$ which is associated with the quantum jump of the system from the exited state $|e\rangle$ to the ground state $|g\rangle$.
By virtue of  Eq.~\eqref{eq:evopardenmat}, the time evolution of the generating function $\rho(s,t)$ is governed by 
\begin{equation}
	\label{eq:evogenfun}
	\frac{d}{dt}\rho(s,t)=\mathcal{C}\rho(s,t)+s\mathcal{J}\rho(s,t).
\end{equation}
By introducing the generalized Bloch vectors defined by
\begin{equation}
	\label{eq:Blovec}
	\begin{split}
		\mathcal{U}(s,t)&=\rho_{eg}(s,t)+\rho_{ge}(s,t),\\
		\mathcal{V}(s,t)&=i\left[\rho_{eg}(s,t)-\rho_{ge}(s,t)\right],\\
		\mathcal{W}(s,t)&=\rho_{ee}(s,t)-\rho_{gg}(s,t),\\
		\mathcal{P}(s,t)&=\rho_{ee}(s,t)+\rho_{gg}(s,t),
		\end{split}
\end{equation}
we can establish the generalized optical Bloch equations describing the transition of the single-molecule system from the exited state $|e\rangle$ to the ground state $|g\rangle$, in matrix form as
\begin{equation}
	\label{eq:Blovecevo}
	\begin{split}
		\frac{d}{dt}\mathbb{Y}(s,t)=\mathcal{M}(s)\mathbb{Y}(s,t),
	\end{split}
\end{equation}
where $\mathbb{Y}(s,t)=\left(\mathcal{U}(s,t),\mathcal{V}(s,t),\mathcal{W}(s,t),\mathcal{P}(s,t)\right)^{\dag}$ is the 4 component vector, and $\mathcal{M}(s)$ is the $4\times4$ coefficient matrix which is given by
\begin{equation}
	\label{eq:coemat}
	\mathcal{M}(s)=\left
	(\begin{array}{cccc}
		\vspace{1.5ex}
		-\dfrac{\Gamma}{2}&-\Delta_{0}&0&0\\
		\vspace{1.5ex}
		\Delta_{0}&-\dfrac{\Gamma}{2}&-\Omega_{0}&0\\
		\vspace{1.5ex}
		0&\Omega_{0}&-\dfrac{\Gamma}{2}(1+s)&-\dfrac{\Gamma}{2}(1+s)\\
		\vspace{1.5ex}
		0&0&-\dfrac{\Gamma}{2}(1-s)&-\dfrac{\Gamma}{2}(1-s)
	\end{array}
	\right).
\end{equation}

Consequently, by setting $s=0$, the probability that the driven two-level single-molecule system has emitted $n$ photons before time $t$ can be defined as
\begin{equation}
	\label{eq:pronpho}
	p_{n}(t)=\frac{1}{n!}\frac{\partial^{n}}{\partial{s^{n}}}\mathcal{P}(s,t)\Big|_{s=0},
\end{equation}
and the $n$-th order factorial moment of the number of emitted photons can be extracted by setting $s=1$ as
\begin{equation}
	\label{eq:kthmom}
	\begin{split}
		\langle N^{(n)}\rangle(t)&=\langle N(N-1)\cdots(N-n+1)\rangle(t)\\
		&=\frac{\partial^{n}}{\partial{s^{n}}}\mathcal{P}(s,t)\Big|_{s=1}.
	\end{split}
\end{equation}
The solution for the $n$-th order derivatives of $\mathcal{P}(s,t)$ with respect to the counting variable $s$ can be achieved by constructing a system of differential equations that includes the generating function $\mathbb{Y}(s,t)$ as well as its derivatives of orders ranging from the 1st to the $n$-th, i.e., $(\partial^{k}/\partial{s^{k}})\mathbb{Y}(s,t)$ for $k=1,2,\ldots,n$ (see Appendix~\ref{sec:appBsolnthord}).

Accordingly, the average waiting time for the first photon emission is expressed as~\cite{PhysRevA78.015402}
\begin{equation}
	\label{eq:waitim}
	\langle\tau\rangle=\int_{0}^{\infty}p_{0}(t)dt,
\end{equation}
the line shape is expressed as the time derivative of the mean photon number
\begin{equation}
	\label{eq:linsha}
		I(t)=\frac{d}{dt}\langle N\rangle(t),
\end{equation}
and the Mandel’s parameter is defined as~\cite{JChemPhys.119.11814,PhysRevA71.033807,JChemPhys.130.244502,JChemPhys.131.214107,JChemPhys.144.064306}
\begin{equation}
	\label{eq:Fanofac}
	Q(t)=\frac{\langle N^{(2)}\rangle(t)-\langle N\rangle^{2}(t)}{\langle N\rangle(t)}.
\end{equation}
The Mandel’s parameter quantifies the statistical fluctuations of photon emission which is negative for sub-Poissonian statistics, zero for Poissonian statistics and positive for super-Poissonian statistics, respectively.

\section{Photon emission under spectral diffusion induced by nonequilibrium environmental fluctuations}
\label{sec:TherFram2}

Under the influence of spectral diffusion, the frequency difference of the single-molecule system fluctuates stochastically as $\omega(t)=\omega_{0}+\xi(t)$, where $\xi(t)$ denotes the environmental noise which is subject to a classical stochastic process.
In the presence of spectral diffusion induced by equilibrium environmental fluctuations, the environmental noise $\xi(t)$ is governed by a stationary stochastic process.~\cite{JChemPhys.128.034106,PhysRevLett.118.140403,JChemPhys.139.134106}
In contrast, when the environmental fluctuations deviate from equilibrium, the environmental noise $\xi(t)$ is subject to a stochastic process with nonstationary statistical properties.~\cite{Kampenbook,JChemPhys.161.044106}

In the following, we will study the photon emission of the driven two-level single-molecule system under the influence of spectral diffusion induced by nonequilibrium environmental fluctuations.
In this situation, the fluctuation term $\xi(t)$ in the frequency difference of the quantum system is governed by a nonstationary stochastic process.~\cite{JChemPhys.161.044106}
The statistical properties are governed by an initial nonstationary distribution $P(\xi,0)$ and its transition probability $P(\xi,t|\xi^{'},t^{'})$, which obeys a master equation
\begin{equation}
	\label{eq:trapro}
	\frac{\partial}{\partial t}P(\xi,t|\xi^{'},t^{'})=\mathcal{Z}P(\xi,t|\xi^{'},t^{'}),
\end{equation}
where $\mathcal{Z}$ is a differential operator only involving the derivatives with respect to $\xi$ for a continuous process, while it is a jump operator for a discrete process.

In the presence of the environmental noise $\xi(t)$, the photon emission of the driven single-molecule system is closely associated with the ensemble-average generating function $\langle\mathbb{Y}(s,t)\rangle$ with 
 $\langle\cdots\rangle$ denoting an average taken over the environmental noise.
Based on Eq.~\eqref{eq:Blovecevo} and within the framework of stochastic Liouville equation,~\cite{PhysRevA105.052443,JChemPhys.161.044106} we can first obtain the time evolution of the marginal average $\overline{\mathbb{Y}}(\xi,s,t)$ as
\begin{equation}
	\label{eq:marave}
	\frac{d}{dt}\overline{\mathbb{Y}}(\xi,s,t)=\mathcal{M}(\xi,s)\overline{\mathbb{Y}}(\xi,s,t)+\mathcal{Z}\overline{\mathbb{Y}}(\xi,s,t),
\end{equation}
with the initial condition $\mathbb{Y}(\xi,s,0)=\langle\mathbb{Y}(s,0)\rangle P(\xi,0)$.
The ensemble-average quantity $\langle\overline{\mathbb{Y}}(s,t)\rangle$ can be derived by the complete average
\begin{equation}
	\label{eq:comave}
	\langle\mathbb{Y}(s,t)\rangle=\int d\xi\overline{\mathbb{Y}}(\xi,s,t),
\end{equation}
where the integral extends over the entire probability distribution of the environmental noise $\xi(t)$ subject to a continuous process, whereas for $\xi(t)$ governed by a noncontinuous process, the integral should be converted into a summation that extends over all discrete states of the environmental noise.

\subsection{Photon emission in the presence of spectral diffusion governed by nonstationary OUN}

We first consider the case that the spectral diffusion induced by nonequilibrium environmental fluctuations is subject to the nonstationary OUN. 
The OUN is a key continuous stochastic process with Gaussian feature which has been widely used to study some important issues related to the dynamics of open quantum systems.~\cite{PhysRevLett.105.015301,PhysRevLett.109.130401,PhysRevA91.022109,JChemPhys.151.164110,PhysRevB97.125435,PhysRevX9.021009,PhysRevB101.174302,PhysRevA103.042607,PhysRevA105.L010601}

For the nonstationary OUN, the differential operator $\mathcal{Z}$ for the transition probability in Eq.~\eqref{eq:trapro} is given by
\begin{equation}
	\label{eq:traproOUN}
	\mathcal{Z}=\gamma\frac{\partial}{\partial \xi}\left(\xi+\sigma^{2}\frac{\partial}{\partial \xi}\right),
\end{equation}
where $\gamma$ is the decay rate which defines the spectral width of the environmental noise, and $\sigma$ denotes the standard deviation of the distribution that describes the strength of the environmental coupling.
The initial condition in the master equation for the transition probability yields $P(\xi,t|\xi^{'},t^{'})=\delta(\xi-\xi^{'})$.

The nonstationary statistical property of the OUN is described by the initial single-point probability~\cite{JChemPhys.161.044106}
\begin{equation}
	\label{eq:inisinpoiOUN}
	P(\xi,0)=\frac{e^{-\frac{(\xi-a\chi)^{2}}{2\sigma^{2}(1-a^{2})}}}{\sqrt{2\pi\sigma^{2}(1-a^{2})}},
\end{equation}
where $\chi$ is an initial parameter and $a$ denotes the nonequilibrium parameter satisfying $|a|<1$.
For the case $a\chi=0$, the OUN exhibits stationary statistics with the variance of fluctuations $\sigma^{2}(1-a^{2})$.
In the case of $a\chi\neq0$, the OUN is nonstationary and the environmental fluctuations are initially out of equilibrium.
As time $t$ evolves, the environmental fluctuations gradually relax to equilibrium with the relaxation rate determined by the decay rate $\gamma$.~\cite{JChemPhys.161.044106}

The complete average of the generating function can be numerically derived as $\langle\mathbb{Y}(s,t)\rangle=\int d\xi\overline{\mathbb{Y}}(\xi,s,t)$ in terms of the time evolution of the marginal average
\begin{equation}
	\label{eq:comaveOUN}
	\begin{split}
	\frac{d}{dt}\overline{\mathbb{Y}}(\xi,s,t)&=\mathcal{M}(\xi,s)\overline{\mathbb{Y}}(\xi,s,t)\\
	&\quad+\gamma\frac{\partial}{\partial \xi}\left(\xi+\sigma^{2}\frac{\partial}{\partial \xi}\right)\overline{\mathbb{Y}}(\xi,s,t),
	\end{split}
\end{equation}
with the initial condition 
\begin{equation}
	\label{eq:iniconOUN}
	\overline{\mathbb{Y}}(\xi,s,0)=\frac{e^{-\frac{(\xi-a\chi)^{2}}{2\sigma^{2}(1-a^{2})}}}{\sqrt{2\pi\sigma^{2}(1-a^{2})}}\langle\mathbb{Y}(s,0)\rangle.
\end{equation}
It can be seen that for the environmental fluctuations in equilibrium ($a=0$), the values 
$\pm\xi$ make equivalent contributions in the complete average $\langle\mathbb{Y}(s,t)\rangle$ whereas for nonequilibrium environmental fluctuations ($a\chi\neq0$), the values symmetrically distributed about $a\chi$ 
contribute equally in the complete average.

\subsection{Photon emission under the influence of spectral diffusion subject to nonstationary RTN}

We now consider the case that the spectral diffusion induced by nonequilibrium environmental fluctuations is governed by the RTN with nonstationary statistics.
The RTN is an important discrete non-Gaussian stochastic process that has been widely used to investigate the environmental influences on open quantum systems, including single-molecule fluorescence, disentanglement, decoherence, and frequency modulation processes.~\cite{NewJPhys.8.1,PhysRevA73.022332,PhysRevB75.054515,JChemPhys.144.024113,PhysRevA100.052104,PhysRevA107.L030601,ApplPhysLett.122.244001}

For the nonstationary RTN, it transits randomly between the values $\pm\nu$ with the 
jump operator $\mathcal{Z}$ for the transition probability in Eq.~\eqref{eq:trapro} given by
\begin{equation}
	\label{eq:traproRTN}
	\mathcal{Z}= \left
	(\begin{array}{cc}
		-\lambda&\lambda\\
		\lambda&-\lambda
	\end{array}
	\right),
\end{equation}
where transition amplitude $\nu$ characterizes the coupling strength of the environmental fluctuations and the switching rate $\lambda$ quantifies the spectral width of the environmental coupling.
The initial condition for the transition probability satisfies $P(\xi,t|\xi^{'},t^{'})=\delta_{\xi,\xi^{'}}$ for $\xi=\pm\nu$.

The nonstationary statistical property of the RTN is determined by the initial single-point probability~\cite{JChemPhys.161.044106}
\begin{equation}
	\label{eq:inisinpoiRTN}
	P(\xi,0)=\frac{1}{2}(1+a)\delta_{\xi,\nu}+\frac{1}{2}(1-a)\delta_{\xi,-\nu},
\end{equation}
with the nonequilibrium parameter $|a|\leq1$.
For the case $a=0$, the RTN displays stationary statistical property.
For $a\neq0$, the RTN exhibits nonstationary statistics with the environmental fluctuations initially in nonequilibrium.
Over time, the environmental fluctuations asymptotically approach equilibrium, with the relaxation time dictated by the switching rate $\lambda$.
The larger the value of $\lambda$, the more rapidly the environmental fluctuations relax to equilibrium; the smaller the value of $\lambda$, the slower the relaxation occurs.~\cite{JChemPhys.161.044106}

The complete average of the generating function can be expressed in an explicit numerical expression as $\langle\mathbb{Y}(s,t)\rangle=\sum_{\pm\nu}\overline{\mathbb{Y}}(\pm\nu,s,t)$ with the time evolution of the marginal average
\begin{equation}
	\label{eq:comaveRTN}
	\begin{split}
		\frac{d}{dt} \left
		(\begin{array}{cc}
			\overline{\mathbb{Y}}(\nu,s,t)\\
			\overline{\mathbb{Y}}(-\nu,s,t)\end{array}\right)&=\mathbb{M}(s)\left
			(\begin{array}{c}
			\overline{\mathbb{Y}}(\nu,s,t)\\
			\overline{\mathbb{Y}}(-\nu,s,t)\end{array}\right),
	\end{split}
\end{equation}
with the coefficient matrix
\begin{equation}
	\label{eq:comaveRTN}
\mathbb{M}(s)=\left
(\begin{array}{cc}
	\mathcal{M}(\nu,s)-\lambda I&\lambda I\\
	\lambda I&\mathcal{M}(-\nu,s)-\lambda I\end{array}\right),
\end{equation}
and the initial condition 
\begin{equation}
	\label{eq:iniconRTN}
	\overline{\mathbb{Y}}(\pm\nu,s,t)=\frac{1}{2}(1\pm a)\langle\mathbb{Y}(s,0)\rangle.
\end{equation}
In the presence of equilibrium environmental fluctuations ($a=0$), the two discrete states  $\pm\nu$ contribute equally to the complete average $\langle\mathbb{Y}(s,t)\rangle$, whereas their contributions are nonequivalent in the complete average for nonequilibrium environmental fluctuations ($a\neq0$).

\section{Results and discussion}
\label{sec:ResuDisc}
In this section, we present the numerical results for the photon counting statistics of the driven single-molecule system under the influence of spectral diffusion arising from nonequilibrium environmental fluctuations.
Generally, in most single-molecule systems (specially the dye molecules), the spontaneous emission rate $\Gamma$ is typically on MHz scale, whereas the frequency difference $\omega_{0}$ is on the GHz-THz scale.~\cite{PhysRevLett.69.1516,AnnuRevPhysChem.48.181,PhysRevLett.83.2722,AnnuRevPhysChem.55.585,RevModPhys.87.1183}
All subsequent analysis is performed under the condition that  $|\Delta_{0}|\ll\omega_{0}+\omega_{L}$,  i.e., in the regime where the RWA holds.
Without the loss of generality, we set the Rabi frequency $\Omega_{0}=\Gamma$ and assume the system is initially prepared in the excited state $|e\rangle$.
Additionally, we choose $\Gamma t=1\times10^{-3}$ as a typical short time scale to ensure that a small number of photons have already been emitted by the single-molecule system.
On such a time scale, the behavior of both the line shape $I(t)$ and Mandel’s parameter $Q(t)$ is very close to that in the steady-state limit under the influence of equilibrium environmental fluctuations.
We analyze in detail how the nonequilibrium characteristics of environmental fluctuations affect photon counting statistics, including the line shape $I(t)$ of emitted photons, and Mandel’s parameter $Q(t)$ which quantifies the statistical fluctuations of the photon emission.
Specifically, we also compare our results with those obtained previously under spectral diffusion induced by equilibrium environmental fluctuations.

\subsection{Photon counting statistics under spectral diffusion induced by nonequilibrium environmental fluctuations subject to nonstationary OUN}
The OUN is a key continuous process exhibiting Gaussian features.
Compared to that with stationary statistical property,  the OUN with nonstationary statistics induces considerable complexity in the dynamical response of the quantum system.
Under the influence of spectral diffusion governed by stationary OUN, the photon emission from the driven single-molecule system has been demonstrated to display distinct statistical characteristics across well-characterized dynamical regimes.~\cite{JChemPhys.121.7914}
In what follows, we  study the influence of the spectral diffusion governed by nonstationary OUN on the statistical properties of the photon emission of the driven two-level single-molecule system.

\subsubsection{Slow modulation limit: $\gamma\ll\sigma$, $\Gamma$, $\Omega_{0}$}
In the slow modulation limit, the decay rate $\gamma$ of the OUN is much smaller than the spontaneous emission rate $\Gamma$, the Rabi frequency $\Omega_{0}$ and the standard deviation $\sigma$. 
In this case, the time required for the environmental fluctuations to relax to equilibrium is exceedingly long, such that the driven single molecule has emitted many photons before the environmental fluctuations reach equilibrium.

\begin{figure}[ht]
	\centering
	\includegraphics[width=3.5in]{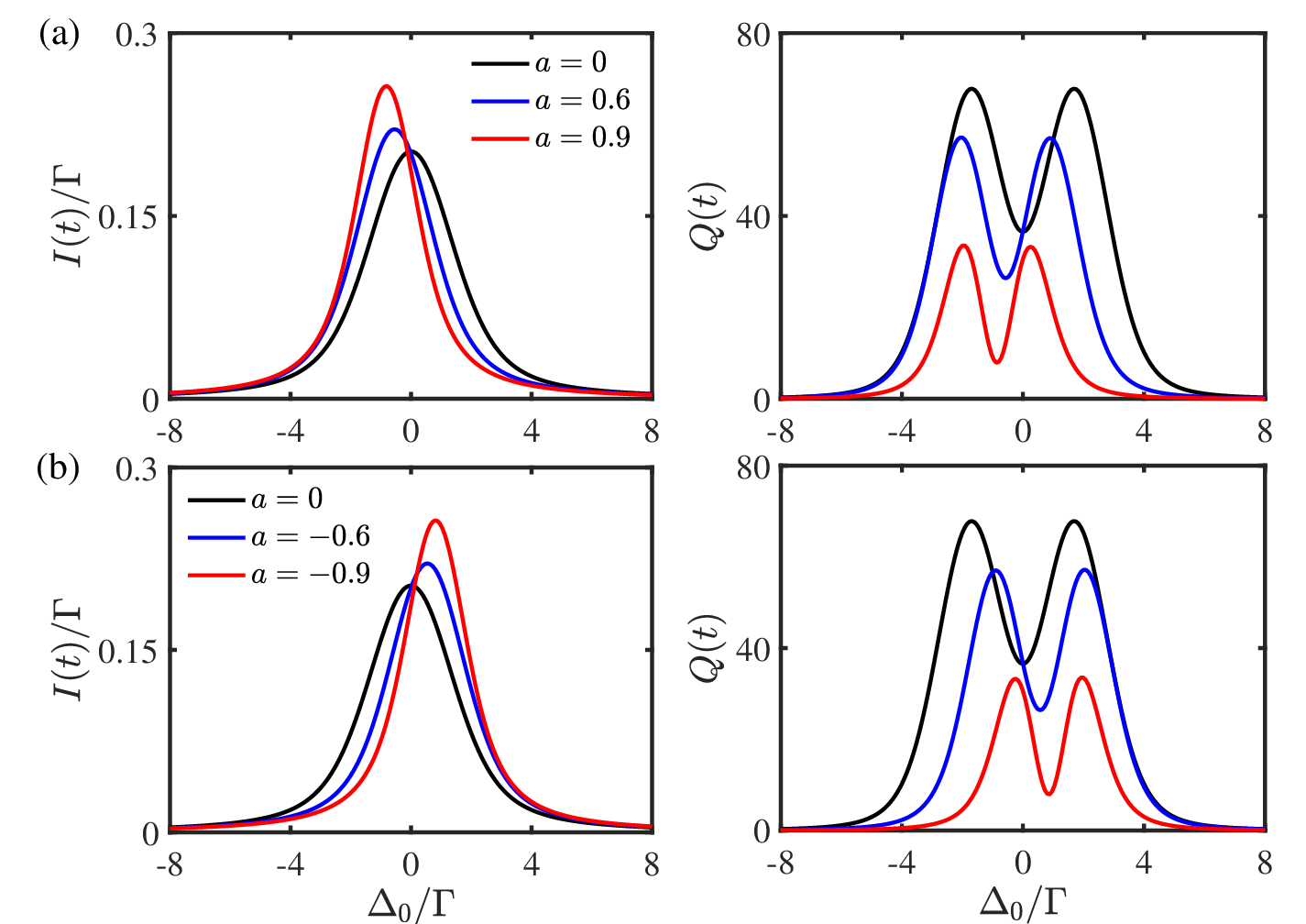}
	\caption{(Color online) The line shape $I(t)$ and Mandel’s parameter $Q(t)$ versus the detuning frequency $\Delta_{0}$ for different nonequilibrium parameter $a$ for a short time $\Gamma t=1\times10^{3}$ in the slow modulation limit (a) for $a>0$ and (b) for $a<0$. Parameters are chosen as: $\Omega_{0}=\Gamma$, $\sigma=\Gamma$, $\chi=\Gamma$ and $\gamma=1\times10^{-4}\Gamma$.}
	\label{fig:slowmoduOUN}
\end{figure}

In Fig.~\ref{fig:slowmoduOUN}, we show the short time line shape $I(t)$ and Mandel’s parameter $Q(t)$ as a function of the detuning frequency $\Delta_{0}$ for different nonequilibrium parameter $a$.
For the equilibrium environmental fluctuations ($a=0$), the line shape $I(t)$ exhibits a single peak centered at zero detuning while the Mandel’s parameter $Q(t)$ displays a symmetric, overlapping double-peak splitting profile centered on $\Delta_{0}=0$. 
For the environmental fluctuations out of equilibrium ($a\chi\neq0$), both the line shape $I(t)$ and Mandel’s parameter $Q(t)$ shift to the negative detuning side for $a>0$ and to the positive detuning side for $a<0$.
As the environmental fluctuations gradually deviate from equilibrium (with the increase of $|a|$), the peak of the line shape $I(t)$ becomes higher and narrower whereas the double peaks of  the Mandel’s parameter $Q(t)$ become shorter and narrower.
These results indicate that at short time scales, the nonequilibrium characteristics of environmental fluctuations can increase the intensity of the line shape, reduce its width, and suppress fluctuations in photon emission.

\begin{figure}[ht]
	\centering
	\includegraphics[width=3.5in]{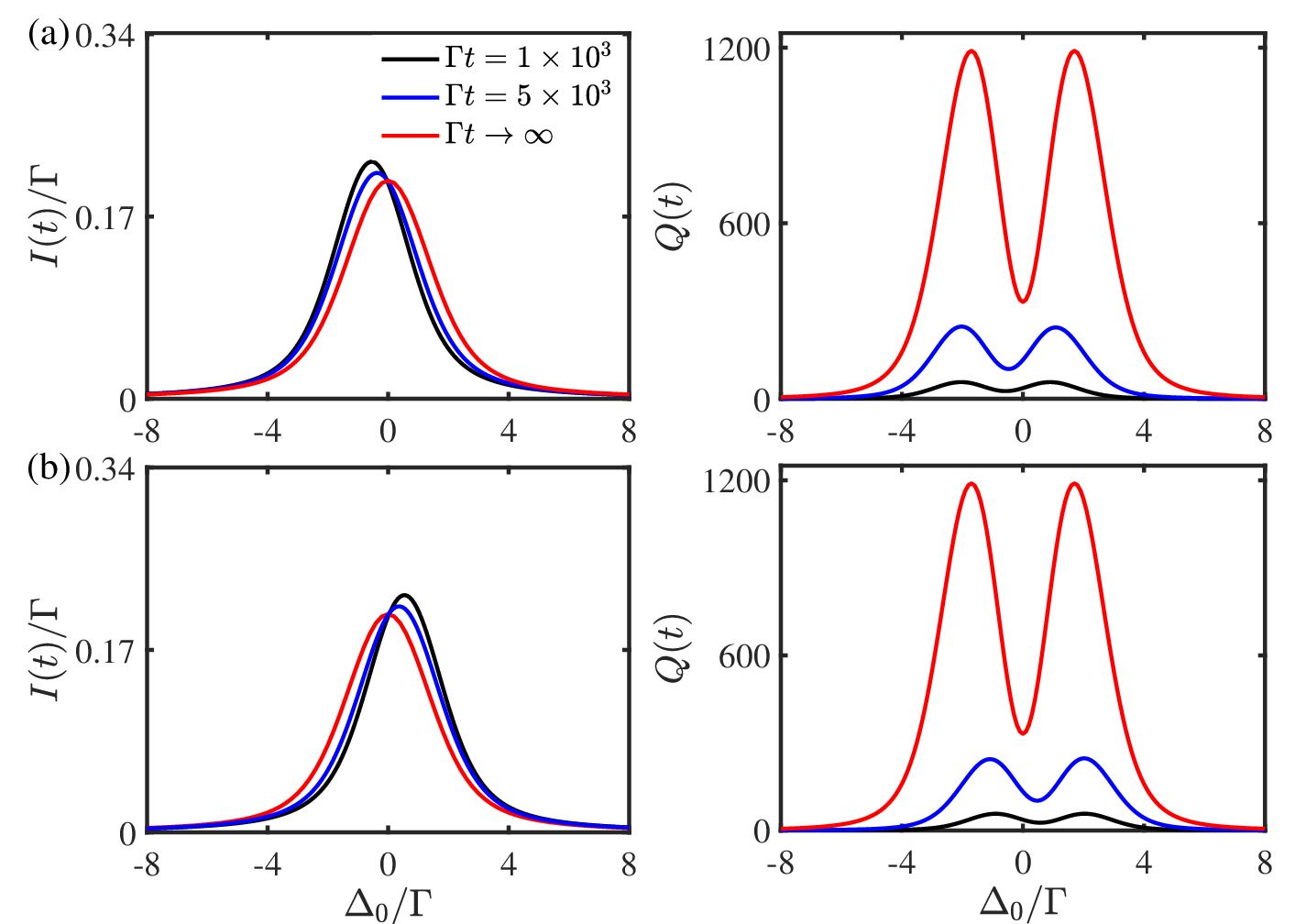}
	\caption{(Color online) The line shape $I(t)$ and Mandel’s parameter $Q(t)$ as a function of the detuning frequency $\Delta_{0}$ for different evolution time $t$ for fixed nonequilibrium parameter $|a|=0.6$ (a) for $a=0.6$ and (b) for $a=-0.6$ in the slow modulation limit. Parameters are set as: $\Omega_{0}=\Gamma$, $\sigma=\Gamma$, $\chi=\Gamma$ and $\gamma=1\times10^{-4}\Gamma$.}
	\label{fig:slowmoduOUNTC}
\end{figure}

To further address the influence of the nonequilibrium characteristics of environmental fluctuations on the photon emission of the driven singe molecule at different time $t$ in the slow modulation limit, we plot the time-dependent line shape $I(t)$ and Mandel’s parameter $Q(t)$ versus the detuning frequency $\Delta_{0}$ for fixed nonequilibrium parameter $|a|=0.6$ from short times to the steady-state limit in Fig.~\ref{fig:slowmoduOUNTC}.
As time $t$ elapses, the symmetry centers of both the line shape $I(t)$ and Mandel’s parameter $Q(t)$ gradually shift from negative detuning to zero detuning for $a>0$ as depicted in Fig.~\ref{fig:slowmoduOUNTC} (a). Conversely, as shown in ~\ref{fig:slowmoduOUNTC} (b) for $a<0$, the symmetry centers of the line shape $I(t)$ and Mandel’s parameter $Q(t)$ exhibit an opposite trend, shifting from positive detuning toward zero detuning.
In the steady state limit, both the line shape $I(t)$ and Mandel’s parameter $Q(t)$ are centered at $\Delta_{0}=0$, in agreement with the existing results obtained in Ref.~\onlinecite{JChemPhys.121.7914}.
This behavior originates from the fact that, at long times, environmental fluctuations relax to equilibrium, and the photon emission process is dominated by the OUN with stationary statistics.

\subsubsection{Fast modulation limit: $\gamma\gg\sigma$, $\Gamma$, $\Omega_{0}$}
\label{sec:fastOUN}
In the fast modulation limit, the decay rate $\gamma$ of the OUN is very large compared with the spontaneous emission rate $\Gamma$, the Rabi frequency $\Omega_{0}$ and the standard deviation $\sigma$. 
In this case, the photon emission of the driven single molecule cannot respond instantaneously to the relaxation of nonequilibrium environmental fluctuations.

\begin{figure}[ht]
	\centering
	\includegraphics[width=3.5in]{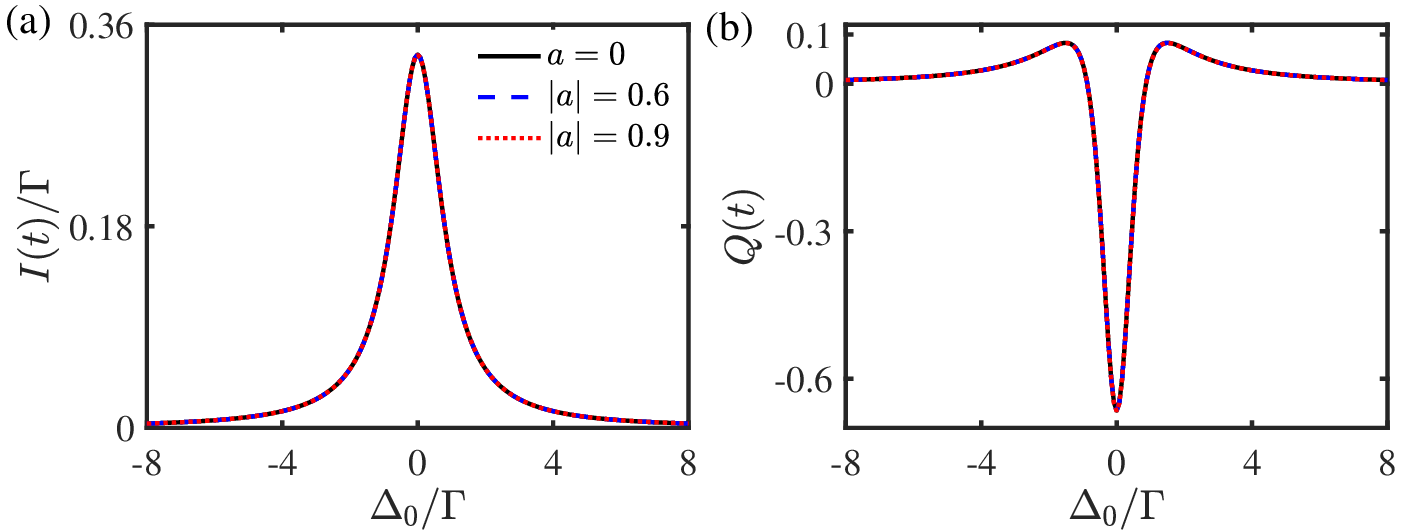}
	\caption{(Color online) The line shape $I(t)$ and Mandel’s parameter $Q(t)$ versus the detuning frequency $\Delta_{0}$ for different nonequilibrium parameter $a$ for a short time $\Gamma t=1\times10^{3}$ in the fast modulation limit (a) for $a>0$ and (b) for $a<0$. Parameters are chosen as: $\Omega_{0}=\Gamma$, $\sigma=\Gamma$, $\chi=\Gamma$ and $\gamma=1\times10^{4}\Gamma$.}
	\label{fig:fastmoduOUN}
\end{figure}

In Fig.~\ref{fig:fastmoduOUN}, we show the line shape $I(t)$ and Mandel’s parameter $Q(t)$ as a function of the detuning frequency $\Delta_{0}$ for different nonequilibrium parameter $a$ at the short-time scale.
The line shape $I(t)$ exhibits a single peak centered at $\Delta_{0}=0$ whereas the Mandel’s parameter $Q(t)$ shows a sharp, symmetric dip that attains a global minimum  $Q_{\mathrm{min}}<0$ at zero detuning. The Mandel's parameter $Q(t)$ is negative in the regime of small detunings and turns positive for large detuning values, which indicates sub-Poissonian and super--Poissonian statistics, respectively.
Obviously, both the line shape $I(t)$ and Mandel’s parameter $Q(t)$ are independent of the nonequilibrium parameter $a$.
This is due to the relaxation of environmental fluctuations to equilibrium preceding the photon emission of the single-molecule system at such a time scale.
It is worth noting that the long time line shape $I(\infty)$ and Mandel’s parameter $Q(\infty)$ will exhibit similar behavior to that at short times, which is consistent with the steady-state results for fast modulation limit obtained in Ref.~\onlinecite{JChemPhys.121.7914}.

\subsection{Photon counting statistics under spectral diffusion arising from  nonequilibrium environmental fluctuations governed by nonstationary RTN}

As a ubiquitous non-Gaussian noise, RTN is typically characterized by discrete switching, and its nonstationary statistical property leads to significant complexity in the dynamical response of the quantum system relative to the case with stationary statistics.
In the presence of spectral diffusion subject to stationary RTN, the photon emission of the driven single-molecule system has been shown to exhibit distinct statistical behaviors across several well-defined dynamical regimes.~\cite{PhysRevLett.93.068302,AnnuRevPhysChem.55.457,JChemPhys.122.184703}
In the following, we study the influence of the spectral diffusion subject to nonstationary RTN on the photon counting statistics of the driven two-level single-molecule system in these dynamical regimes, respectively.

\subsubsection{Slow modulation limit: $\lambda\ll\nu$, $\Gamma$, $\Omega_{0}$}

In the slow modulation limit, the switching rate $\lambda$ of the RTN is very small compared with the spontaneous emission rate $\Gamma$, the Rabi frequency $\Omega_{0}$, and the transition amplitude $\nu$.~\cite{JChemPhys.121.7914,JChemPhys.122.184703}
In this case, the time required for the single-molecule system to transit between the two states $\omega_{0}\pm\nu$ and the relaxation time of the nonequilibrium environmental fluctuations, are both extremely long, such that the system emits numerous photons between successive transitions and prior to the environmental fluctuations reaching equilibrium.

\paragraph{Slow modulation limit strong coupling regime.}

\begin{figure}[ht]
	\centering
	\includegraphics[width=3.5in]{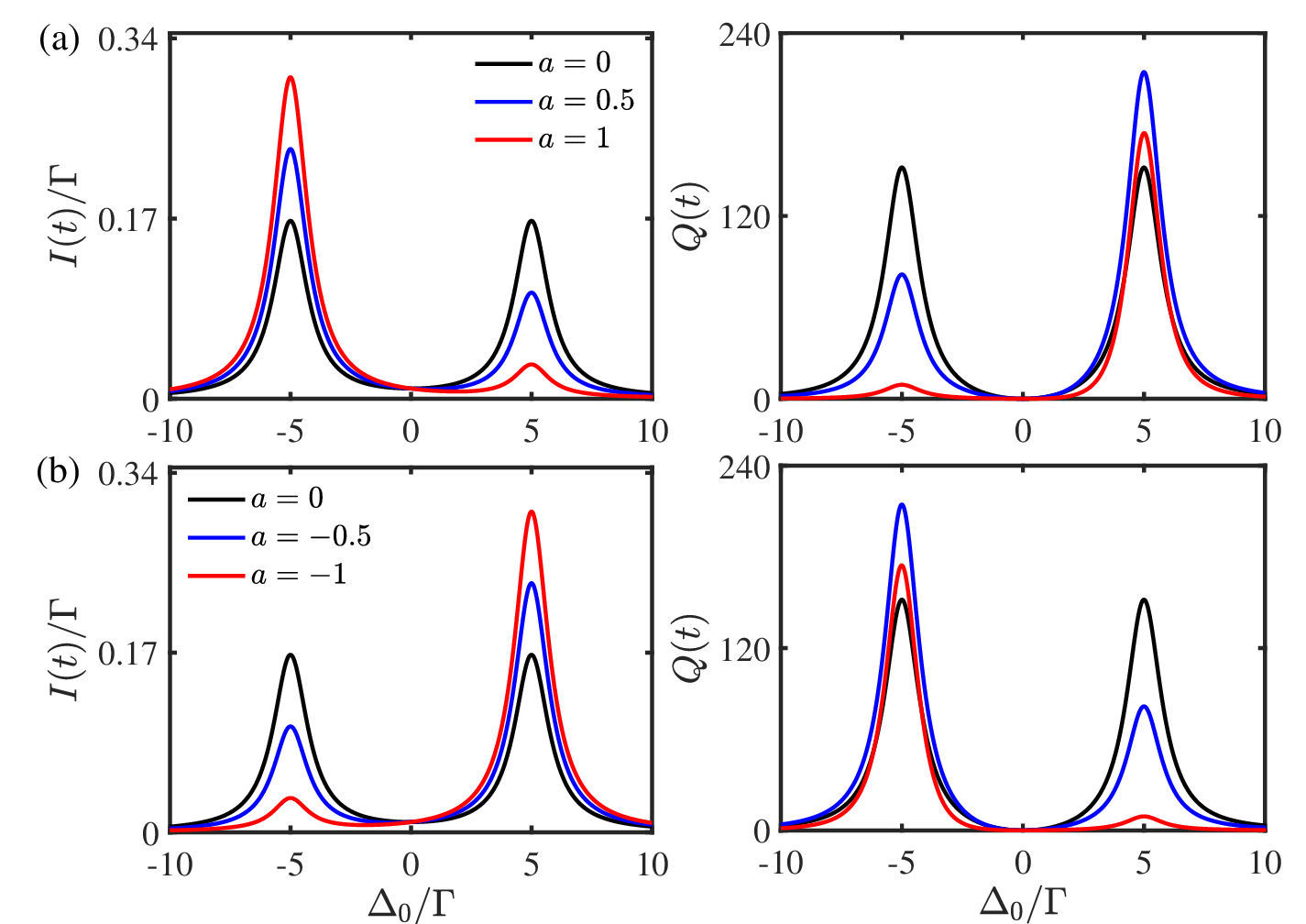}
	\caption{(Color online) The line shape $I(t)$ and Mandel’s parameter $Q(t)$ versus the detuning frequency $\Delta_{0}$ for different nonequilibrium parameter $a$ for a short time $\Gamma t=1\times10^{3}$ in the slow modulation limit strong coupling regime (a) for $a>0$ and (b) for $a<0$. Parameters are chosen as: $\Omega_{0}=\Gamma$, $\nu=5\Gamma$ and $\lambda=1\times10^{-4}\Gamma$.}
	\label{fig:RTNslowmoduSC}
\end{figure}

We first display the short time line shape $I(t)$ and Mandel’s parameter $Q(t)$ as a function of the detuning frequency $\Delta_{0}$ for varying values of the nonequilibrium parameter $a$ in the slow modulation limit strong coupling regime in Fig.~\ref{fig:RTNslowmoduSC}.
Both the line shape $I(t)$ and Mandel’s parameter $Q(t)$ exhibit splitting behavior with two well separated peaks centered at $\Delta_{0}=\pm\nu$.
When the nonequilibrium parameter $a=0$, the two split peaks of both the line shape $I(t)$ and Mandel’s parameter $Q(t)$ are symmetric. 
In contrast, when the nonequilibrium parameter $a\neq0$, the line shape $I(t)$ and Mandel’s parameter $Q(t)$ display asymmetric profiles, each with one dominant main peak and one weaker secondary peak.
The positions of the main peaks of the line shape $I(t)$ and Mandel’s parameter $Q(t)$ are determined by the sign of $a$.
As depicted in Fig.~\ref{fig:RTNslowmoduSC} (a), when $a>0$, the main peak of the line shape $I(t)$ lies on the negative detuning $\Delta_{0}=-\nu$ while the main peak of the Mandel’s parameter $Q(t)$ is located on the positive detuning $\Delta_{0}=\nu$.
When $a<0$, as shown in Fig.~\ref{fig:RTNslowmoduSC} (b), the main peak of the line shape I is on the positive detuning side at $\Delta_{0}=\nu$, whereas the main peak of Mandel’s parameter $Q(t)$ is on the negative detuning side at $\Delta_{0}=-\nu$.
Additionally, the larger the absolute value of $a$, the more pronounced the asymmetric behavior in the line shape $I(t)$ and Mandel’s parameter $Q(t)$ becomes.
As the environmental fluctuations gradually depart from equilibrium ($|a|$ increases from 0 to 1), the main peak of the line shape $I(t)$ rises in height and undergoes broadening, whereas the secondary peak decreases in height and narrows in width.
In contrast, with the increase of $|a|$, the secondary peak of the Mandel’s parameter $Q(t)$ increases in height and broadens, while the main peak first broadens and then narrows, with its height exhibiting a trend of first increasing and then decreasing.
These results indicate that the nonequilibrium characteristics of environmental fluctuations can lead to asymmetric behavior in line shape $I(t)$ and Mandel’s parameter $Q(t)$, and these phenomena occur for short times in the slow modulation limit strong coupling regime.

\begin{figure}[ht]
	\centering
	\includegraphics[width=3.5in]{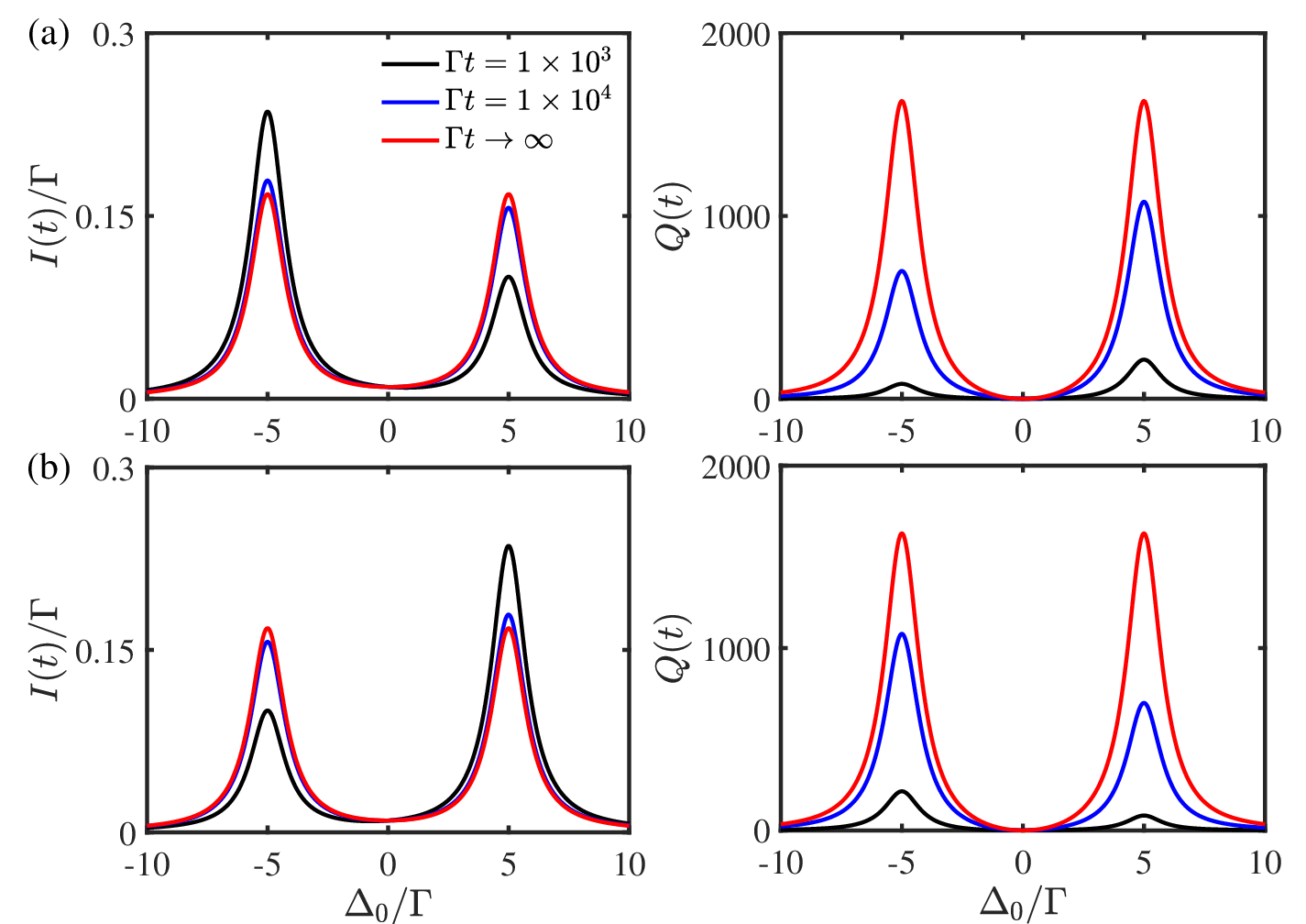}
	\caption{(Color online) The line shape $I(t)$ and Mandel’s parameter $Q(t)$ as a function of the detuning frequency $\Delta_{0}$ for different evolution time $t$ for fixed nonequilibrium parameter $|a|=0.5$ (a) for $a=0.5$ and (b) for $a=-0.5$ in the slow modulation limit strong coupling regime. Parameters are set as: $\Omega_{0}=\Gamma$, $\nu=5\Gamma$, and $\lambda=1\times10^{-4}\Gamma$.}
	\label{fig:RTNslowmoduSCTC}
\end{figure}

To further study the time evolution of $I(t)$ and $Q(t)$ from short times to the steady-state limit  in the slow modulation limit strong coupling regime, we plot the line shape $I(t)$ and Mandel’s parameter $Q(t)$ versus the detuning frequency $\Delta_{0}$ for fixed nonequilibrium parameter $|a|=0.5$ for different evolution time $t$ in Fig.~\ref{fig:RTNslowmoduSCTC}.
As time $t$ evolves, the main peak of the line shape $I(t)$ decreases in height and undergoes narrowing behavior, whereas the secondary peak increases in height and displays broadening behavior.
In contrast, both the main and secondary peaks of the Mandel’s $Q(t)$ increase in height and undergo broadening as the evolution time $t$ increases.
In the long time limit, both the line shape $I(\infty)$ and Mandel’s parameter $Q(\infty)$ display  symmetric profiles with two peaks of equal height centered at $\Delta_{0}=\pm\nu$. 
These trends result from the fact that the environmental fluctuations relax to equilibrium in the long time limit.

\paragraph{Slow modulation limit intermediate coupling regime.}

\begin{figure}[ht]
	\centering
	\includegraphics[width=3.5in]{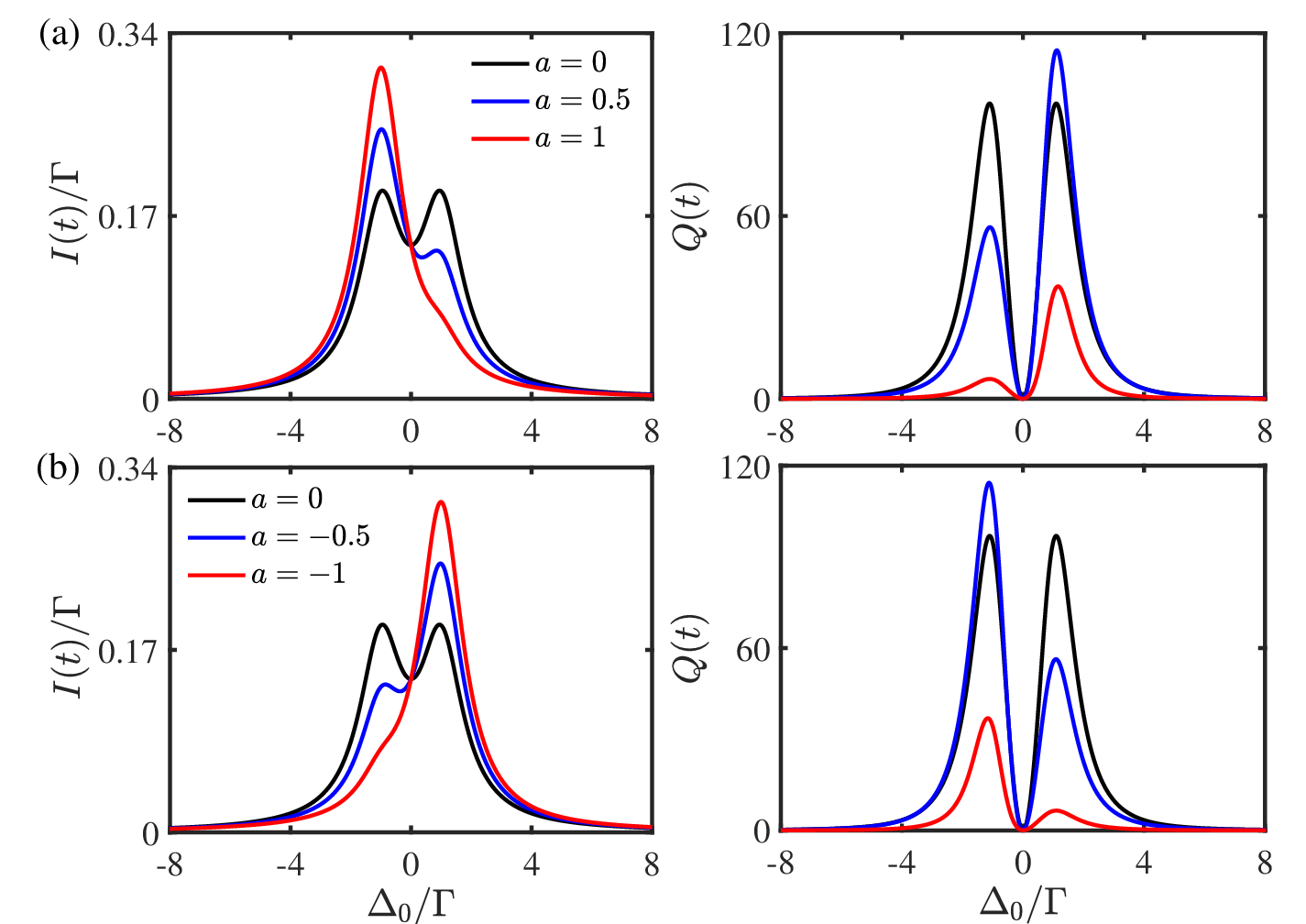}
	\caption{(Color online) The line shape $I(t)$ and Mandel’s parameter $Q(t)$ for different nonequilibrium parameter $a$ versus the detuning frequency $\Delta_{0}$ at a short time  $\Gamma t=1\times10^{3}$ in the slow modulation limit intermediate coupling regime  (a) for $a>0$ and (b) for $a<0$. Parameters are chosen as: $\Omega_{0}=\Gamma$, $\nu=\Gamma$ and $\lambda=1\times10^{-4}\Gamma$.}
	\label{fig:RTNslowmoduIC}
\end{figure}

We next plot the line shape $I(t)$ and Mandel’s parameter $Q(t)$ at the short-time scale as a function of the detuning frequency $\Delta_{0}$ for different values of the nonequilibrium parameter $a$ in the slow modulation limit intermediate coupling regime.
As shown in Fig.~\ref{fig:RTNslowmoduIC}, the line shape $I(t)$ exhibits a partially overlapping double-peak structure while the Mandel’s parameter $Q(t)$ splits into two well separated peaks and attains its minimum when the detuning is zero.
For the nonequilibrium parameter $a=0$, the line shape $I(t)$ and Mandel’s parameter $Q(t)$ display symmetric behavior with two peaks of equal height, whereas they both exhibit asymmetric behavior when the nonequilibrium parameter $a\neq0$.
As the environmental fluctuations gradually deviate from equilibrium ($|a|$ increases from 0 to 1), the symmetric behavior in the line shape $I(t)$ and Mandel’s parameter $Q(t)$ grows more and more obvious.
The line shape $I(t)$ becomes concentrated on one side of the double-peak structure, with increased peak height and broadened width.
Correspondingly, the Mandel’s parameter $Q(t)$ decreases on the side the line shape $I(t)$ concentrated while it first increases and then decreases on the other side.

\begin{figure}[ht]
	\centering
	\includegraphics[width=3.5in]{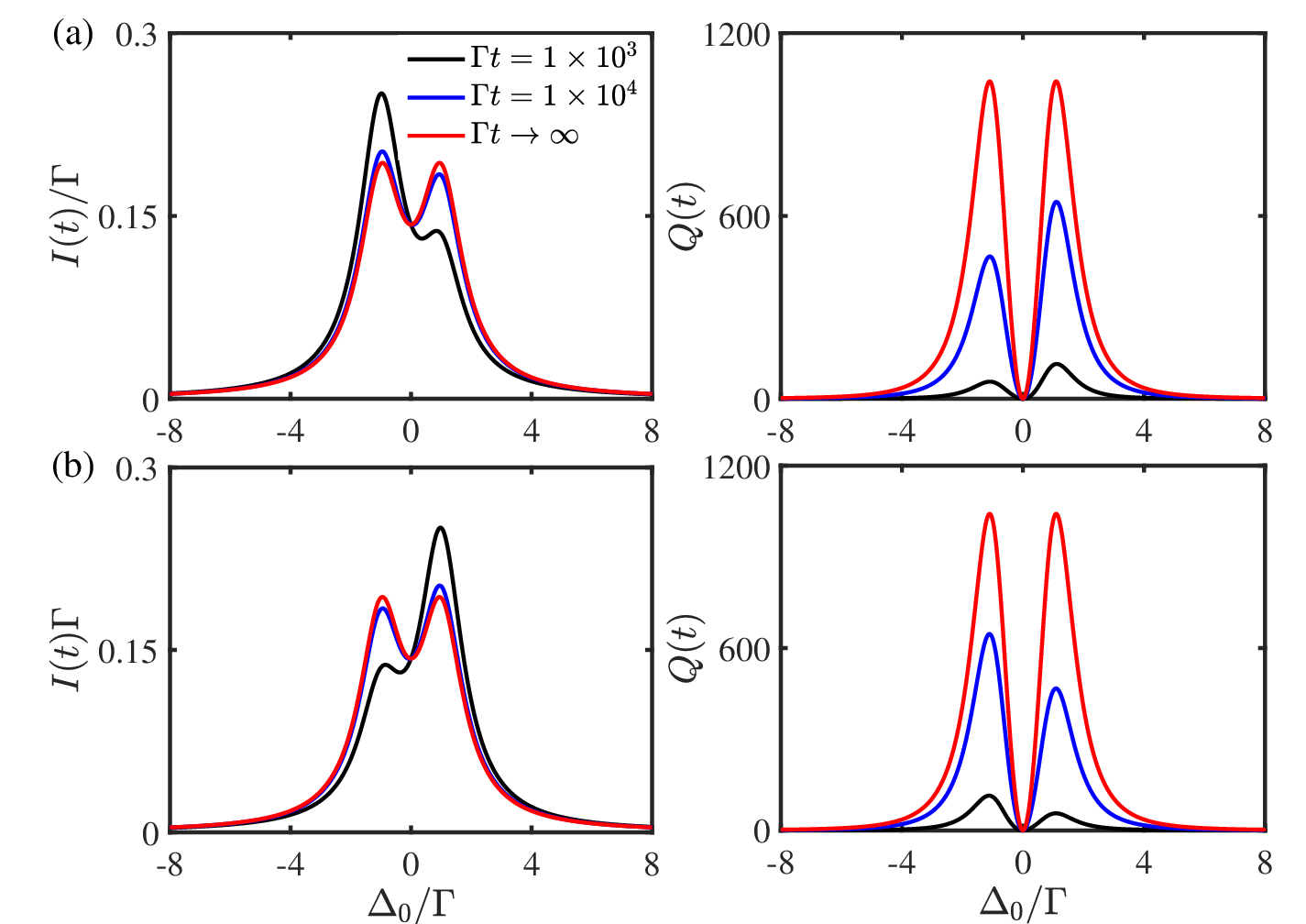}
	\caption{(Color online) The line shape $I(t)$ and Mandel’s parameter $Q(t)$ versus the detuning frequency $\Delta_{0}$ for different evolution time for fixed nonequilibrium parameter $|a|=0.5$ in the slow modulation regime and intermediate coupling regime (a) for $a=0.5$ and (b) for $a=-0.5$. Parameters are chosen as: $\Omega_{0}=\Gamma$, $\nu=\Gamma$ and $\lambda=1\times10^{-4}\Gamma$.}
	\label{fig:RTNslowmoduICTC}
\end{figure}

 We further plot the line shape $I(t)$ and Mandel’s parameter $Q(t)$ as a function of the detuning frequency $\Delta_{0}$ for fixed nonequilibrium parameter $|a|=0.5$ for different evolution time from short times to the long time limit in the slow modulation limit intermediate coupling regime in Fig.~\ref{fig:RTNslowmoduICTC}.
As time $t$ evolves, the asymmetric behavior in the line shape $I(t)$ and Mandel’s parameter $Q(t)$ becomes increasingly less obvious and vanishes in the steady state.
In addition, for the line shape $I(t)$ with a partially overlapping double-peak structure, the major peak exhibits a noticeable decrease in both height and width, while the minor peak shows a significant increase in both its height and width simultaneously.
In contrast, with the increase of evolution time $t$, both the major and minor peaks of Mandel’s parameter $Q(t)$ exhibit an increase in height accompanied by distinct broadening.

\begin{figure}[ht]
	\centering
	\includegraphics[width=3.5in]{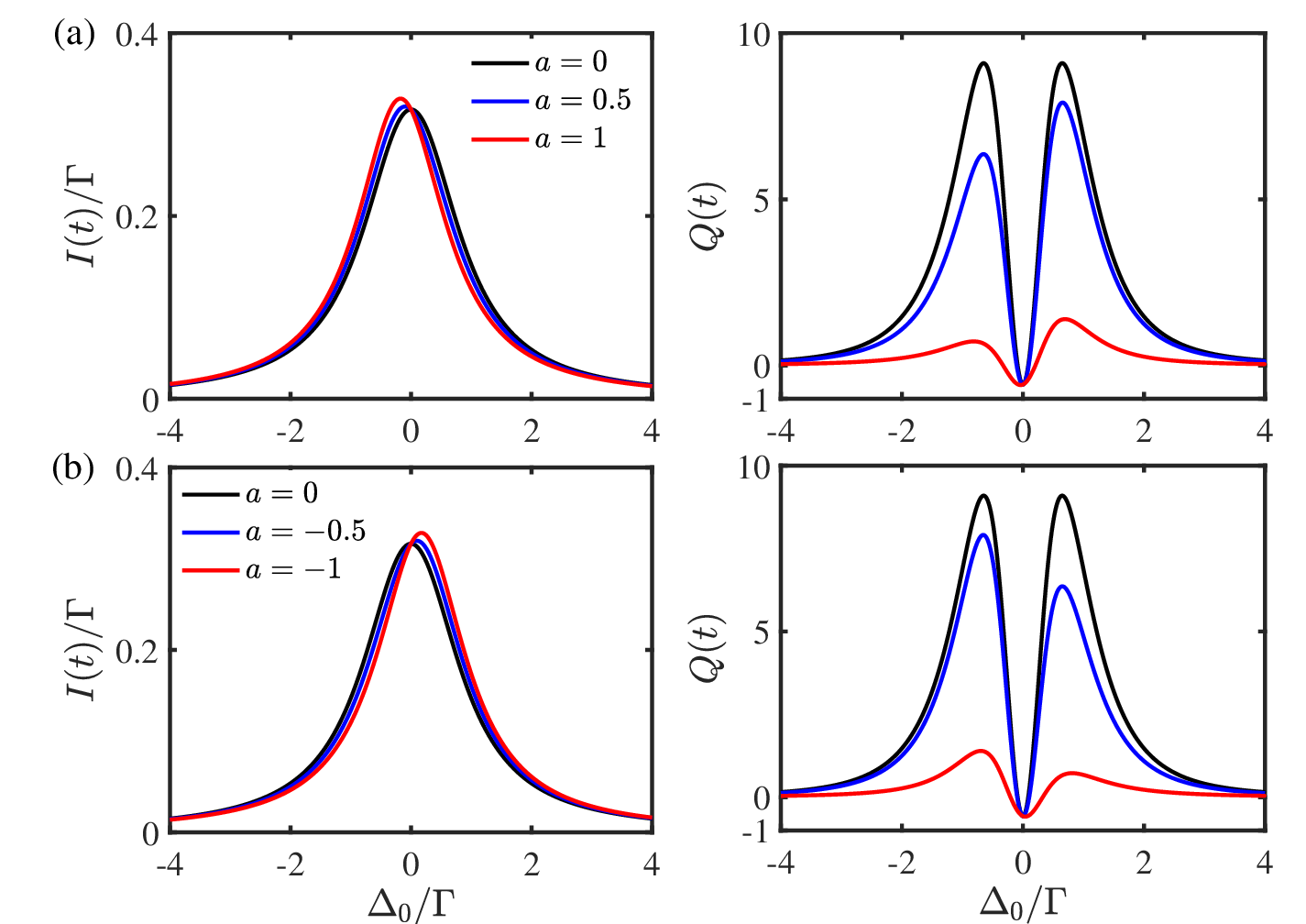}
	\caption{(Color online) The line shape $I(t)$ and Mandel’s parameter $Q(t)$ for different nonequilibrium parameter $a$ versus the detuning frequency $\Delta_{0}$ for short times  $\Gamma t=1\times10^{3}$ in the slow modulation limit weak coupling regime  (a) for $a>0$ and (b) for $a<0$. Parameters are chosen as: $\Omega_{0}=\Gamma$, $\nu=0.2\Gamma$ and $\lambda=1\times10^{-4}\Gamma$.}
	\label{fig:RTNslowmoduWC}
\end{figure}

\paragraph{Slow modulation limit weak coupling regime.}
We now illustrate the short time line shape $I(t)$ and Mandel’s parameter $Q(t)$ varying with the detuning frequency $\Delta_{0}$ for distinct values of the nonequilibrium parameter $a$ in the slow modulation limit weak coupling regime in Fig.~\ref{fig:RTNslowmoduWC}.
When the nonequilibrium parameter $a=0$, the line shape $I(t)$ exhibits a single peak centered at $\Delta_{0}=0$ while the Mandel’s parameter $Q(t)$ shows a splitting into two symmetric peaks and a minimum at zero detuning with $Q_{\mathrm{min}<0}$ which indicates the sub-Poissonian statistics of the photon emission.
As the environmental fluctuations gradually depart from equilibrium (with $|a|$ increasing  from 0 to 1), both the line shape $I(t)$ and Mandel’s parameter $Q(t)$ begin to show asymmetric behavior.
In addition, the height of the peak of line shape $I(t)$ increases slightly; for $a>0$, the center of the peak shifts toward negative detuning, while for $a<0$, it shifts toward positive detuning.
Furthermore, the Mandel’s parameter $Q(t)$ gradually decreases and narrows in width; for $a>0$, the decrease on the negative detuning side is significantly faster than that on the positive detuning side, while for $a<0$, the decrease on the positive detuning side is markedly faster than that on the negative detuning side.

\begin{figure}[ht]
	\centering
	\includegraphics[width=3.5in]{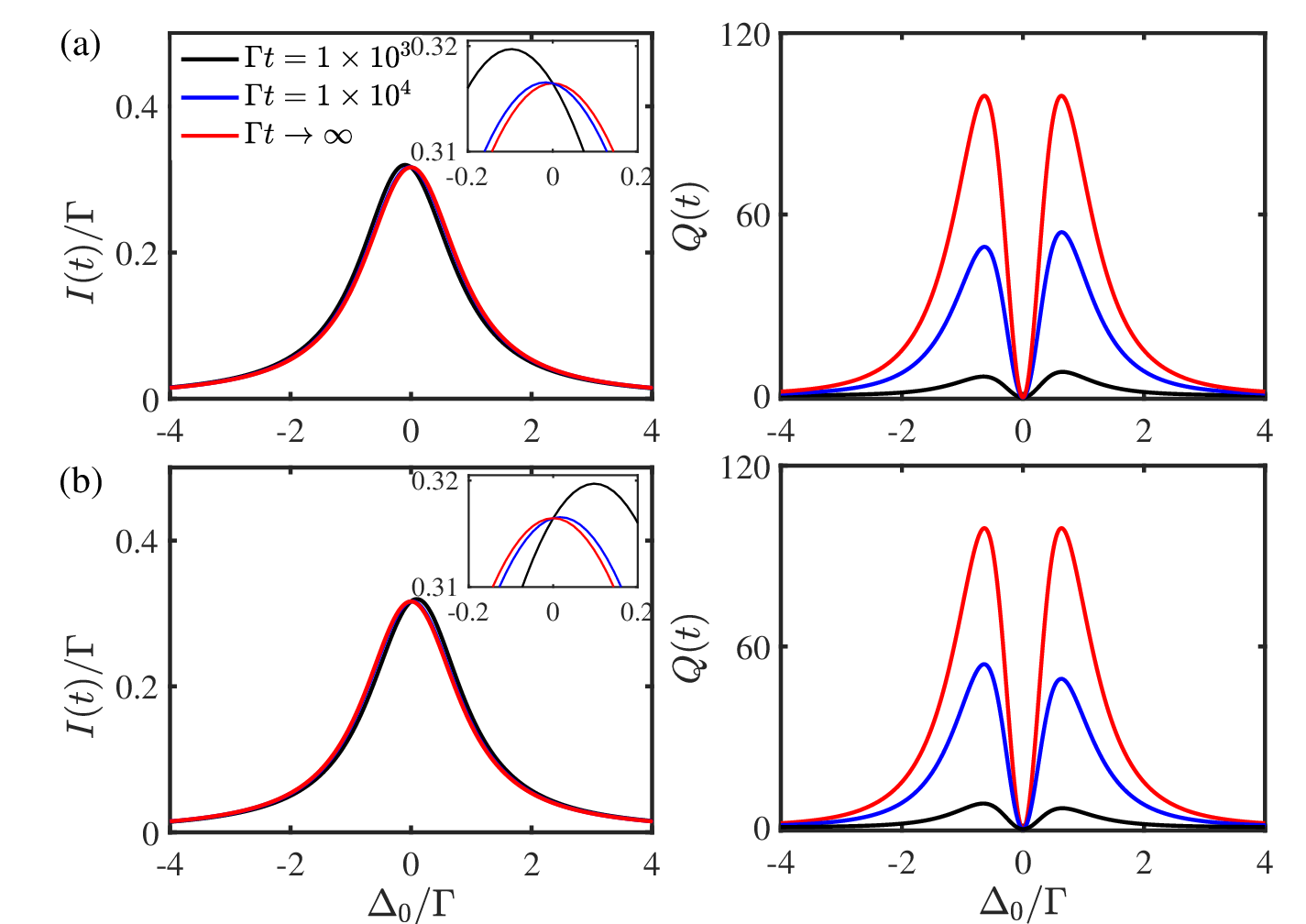}
	\caption{(Color online) The line shape $I(t)$ and Mandel’s parameter $Q(t)$ versus the detuning frequency $\Delta_{0}$ for varying evolution times for fixed nonequilibrium parameter $|a|=0.5$ in the slow modulation limit weak coupling regime (a) for $a=0.5$ and (b) for $a=-0.5$. Parameters are chosen as: $\Omega_{0}=\Gamma$, $\nu=0.2\Gamma$ and $\lambda=1\times10^{-4}\Gamma$.}
	\label{fig:RTNslowmoduWCTC}.
\end{figure}

 We also plot the line shape $I(t)$ and Mandel’s parameter $Q(t)$ as a function of the detuning frequency $\Delta_{0}$ for fixed nonequilibrium parameter $|a|=0.5$ for different evolution time from short times to the long time limit in the slow modulation limit weak coupling regime in Fig.~\ref{fig:RTNslowmoduWCTC}.
As time $t$ elapses, the asymmetric profiles of the line shape $I(t)$ and Mandel’s parameter $Q(t)$ observed at short timescales gradually become symmetric in the long time limit.
For $a>0$, the center of the peak of the line shape $I(t)$ shifts from the negative detuning to zero detuning, while for $a<0$, it shifts from positive detuning to zero detuning.
The Mandel’s parameter $Q(t)$ increases and exhibits broadening as time evolves; for $a>0$, 
the increase on the positive detuning side is markedly faster than that on the negative detuning side, while for $a<0$, the increase on the negative detuning side is significantly faster than that on the positive detuning side.

\paragraph{Brief summary.}
In the slow modulation limit, both the line shape $I(t)$ and Mandel's parameter $Q(t)$ show asymmetric behavior as a function of the detuning frequency $\Delta_{0}$ at the short-time scale when the environmental fluctuations are in nonequilibrium ($a\neq0$).
The steady-state line shape $I(t)$ and Mandel's parameter $Q(t)$ display symmetric behavior, a direct consequence of environmental fluctuations relaxing to equilibrium in the long time limit, which are consistent with the existing results in Refs.~\onlinecite{PhysRevLett.93.068302,JChemPhys.122.184703}.
Notably, in the limiting case $\lambda\rightarrow0$ as adopted in Ref.~\onlinecite{JChemPhys.122.184703}, the asymmetric behavior in both the line shape $I(t)$ and Mandel’s parameter $Q(t)$ persists indefinitely even in the long time limit.
This is attributed to the fact that environmental fluctuations persist in nonequilibrium  throughout and are unable to relax to equilibrium under such a limiting condition.

\subsubsection{Fast modulation limit: $\lambda\gg\nu$,  $\Gamma$, $\Omega_{0}$}
In the fast modulation limit, the switching rate $\lambda$ of the RTN is much larger than the spontaneous emission rate $\Gamma$, the Rabi frequency $\Omega_{0}$ and the transition amplitude $\nu$. 
In this case, the single-molecule system transits rapidly between the two states $\omega_{0}\pm\nu$, and the environmental fluctuations relax to equilibrium in an extremely short time, such that the system has not emitted photons before the environmental fluctuations reach equilibrium.

\begin{figure}[ht]
	\centering
	\includegraphics[width=3.5in]{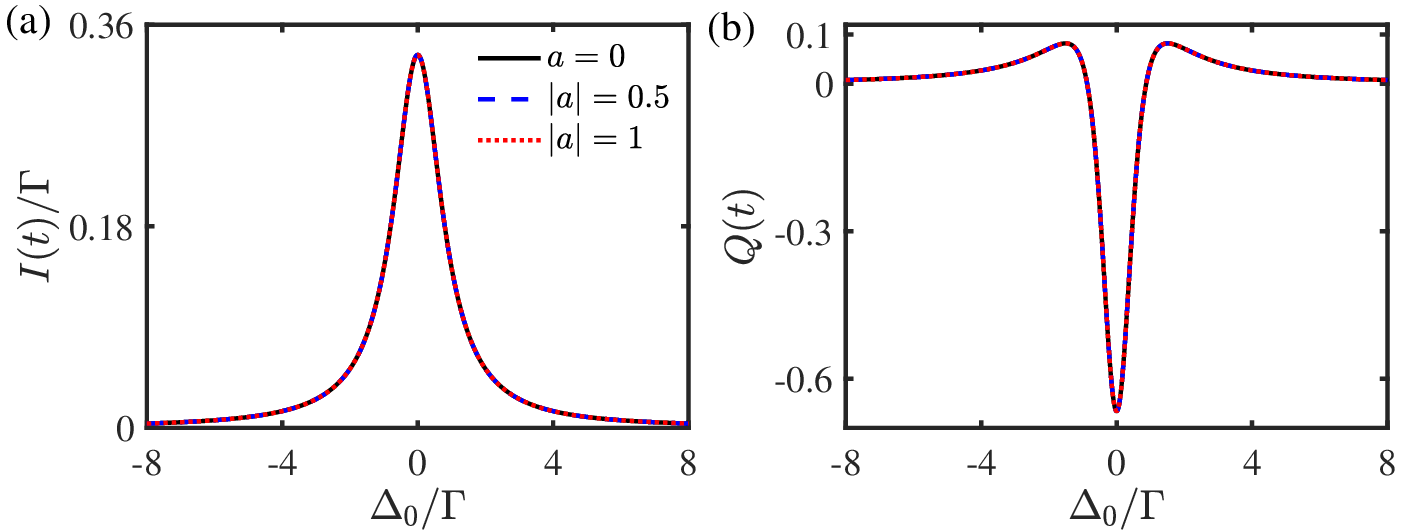}
	\caption{(Color online) The line shape $I(t)$ and Mandel’s parameter $Q(t)$ versus the detuning frequency $\Delta_{0}$ for different nonequilibrium parameter $a$ for a short time $\Gamma t=1\times10^{3}$ in the fast modulation limit strong coupling regime (a) for $a>0$ and (b) for $a<0$. Parameters are chosen as: $\Omega_{0}=\Gamma$, $\nu=5\Gamma$ and $\lambda=1\times10^{-4}\Gamma$.}
	\label{fig:fastmoduRTN}
\end{figure}

In Fig.~\ref{fig:fastmoduRTN}, we show the line shape $I(t)$ and Mandel’s parameter $Q(t)$ as a function of the detuning frequency $\Delta_{0}$ for different nonequilibrium parameter $a$ for a short time scale $\Gamma t=1\times10^{3}$. 
Both the line shape $I(t)$ and Mandel’s parameter $Q(t)$ display similar behavior as that induced by the OUN described in Sec.~\ref{sec:fastOUN} due to the fact that the RTN recovers one with Gaussian features in the limiting case $\lambda\gg\nu$.~\cite{JStatPhys.31.467}
Additionally, neither the line shape $I(t)$ nor Mandel’s parameter $Q(t)$ depends on the nonequilibrium parameter $a$ owing to the rapid relaxation of nonequilibrium environmental fluctuations in the fast modulation limit.
It is also worth mentioning that the steady-state line shape $I(\infty)$ and Mandel’s parameter $Q(\infty)$ will display similar behavior to that at short times, in agreement with the existing results obtained in the long time limit in Refs.~\onlinecite{PhysRevLett.93.068302,JChemPhys.122.184703}.

\section{Conclusions and outlook}
\label{sec:Conc}
In this paper, we have theoretically studied the photon emission of a driven two-level single-molecule system in the presence of spectral diffusion induced by nonequilibrium environmental fluctuations. 
Using the generating function method and the stochastic Liouville equation, we analyze the influence of environmental nonequilibrium characteristics on the intensity and statistical fluctuations of photon emission from the single-molecule system in the presence of OUN and RTN with nonstationary statistics.
In the slow modulation limit, the nonequilibrium characteristics of environmental fluctuations play an essential role in photon counting statistics at short time scales.
More specifically, in the presence of nonstationary OUN, the nonequilibrium environmental fluctuations give rise to shifts in the symmetry centers of both the line shape and Mandel’s parameter while they lead to the symmetric behavior in the line shape and Mandel’s parameter under the influence of nonstationary RTN.
In contrast, the environmental nonequilibrium characteristics exert no discernible impact on the statistical properties of photon emission in the fast modulation limit or in the long time limit, as the environmental fluctuations have already relaxed to equilibrium.

In practice, we shall also take into account the temporal resolution in single-photon measurements.
With sufficiently fine photon-counting time resolution, the influence of the nonequilibrium characteristics of environmental fluctuations can in principle be resolved.
Both slow and fast relaxations of nonequilibrium environmental fluctuations processes are not significantly affected by typical temporal resolution limits.
Furthermore, employing properly optimized conditions for photon counting can further improve the sensitivity to environmental relaxation processes in future experiments.~\cite{JPhysChemB110.19040}
Our present findings reveal the mechanisms by which the nonequilibrium environmental fluctuations influence the statistical properties associated with photon emission in single-molecule systems, thereby offering a theoretical and experimental foundation for discriminating between the equilibrium and nonequilibrium characteristics of environmental fluctuations. 

\begin{acknowledgments}
  This work was supported by the National Natural Science Foundation of China (Grant Nos. 12174221 and 12234013).
\end{acknowledgments}

\section*{Author declarations}
\subsection*{Conflict of Interest}
The authors have no conflicts to disclose.
\subsection*{Author Contributions}
\textbf{Xiangji Cai:} Conceptualization (equal); Formal analysis (equal); Investigation (lead); Writing – original draft (lead); Writing – review \& editing (equal). 
\textbf{Yonggang Peng:} Conceptualization (equal); Formal analysis (equal); Investigation (supporting); Writing – review \& editing (equal).
\textbf{Yujun Zheng:} Conceptualization (equal); Formal analysis (equal); Investigation (supporting); Supervision (lead); Writing – review \& editing (equal).

\section*{Data Availability}
The data that support the findings of this study are available from the authors upon reasonable request.

\appendix
\section{Derivation of the master equation~\eqref{eq:evol} for the driven single-molecule system}
\label{sec:appAmasequ}
Since the size of the single-molecule system is much smaller than the wavelength of the laser field, the interaction between the system and the laser field can be treated within the dipole approximation.
Under this approximation, the driven single-molecule system is described by a time-dependent Hamiltonian~\cite{Cohen-Tannoudjibook}
\begin{equation}
	\label{eq:Hamitimdep1}
	\mathcal{H}(t)=\frac{\hbar}{2}\omega_{0}\sigma_{z}-\bm{d}\cdot E\cos\left(\omega_{L}t\right),
\end{equation}
where $\bm{d}=d_{eg}\sigma_{x}$ denotes the electric dipole operator with the matrix element $d_{eg}=\langle e|\bm{d}|g\rangle$, and $E$ is the amplitude of the laser field.

By phenomenologically adding the radiative decay induced by the vacuum fluctuations of the quantum radiation field initially at zero temperature, the time evolution of the driven single-molecule system satisfies the quantum master equation~\cite{Cohen-Tannoudjibook}
\begin{equation}
	\label{eq:evolappA}
	\frac{d}{dt}\varrho(t)=-\frac{i}{\hbar}[\mathcal{H}(t),\varrho(t)]+\Gamma\bigg[\sigma_{-}\varrho(t)\sigma_{+}-\frac{1}{2}\{\sigma_{+}\sigma_{-},\varrho(t)\}\bigg].
\end{equation}
Here, the modifications of spontaneous emission in the presence of the incident radiation are neglected, which is valid if the effect of the coupling with this radiation can be disregarded over the correlation time of the vacuum fluctuations responsible for spontaneous emission.  

Under the near-resonance condition, i.e., when $|\omega_{0}-\omega_{L}|\ll\omega_{0}+\omega_{L}$,
the time-dependent Hamiltonian in Eq.~\eqref{eq:Hamitimdep1} can be rewritten within the RWA as
\begin{equation}
	\label{eq:Hamitimdep2}
	\mathcal{H}_{\mathrm{RWA}}(t)=\frac{\hbar}{2}\omega_{0}\sigma_{z}+\frac{\hbar}{2}\Omega_{0}\left(\sigma_{+}e^{-i\omega_{L}t}+\sigma_{-}e^{i\omega_{L}t}\right),
\end{equation}
where $\Omega_{0}=-d_{eg}\cdot E/\hbar$ stands for the Rabi frequency.
 Under the RWA, we retain only the resonant processes in which the single-molecule system is driven from the ground state $|g\rangle$ to excited state $|e\rangle$ by absorbing a photon, or from the excited state $|e\rangle$ to ground state $|g\rangle$ by emitting a photon. Conversely, we neglect the counter-rotating processes, where the system is driven from he excited state $|e\rangle$ to ground state $|g\rangle$ by absorbing a photon, or from the ground state $|g\rangle$ to excited state $|e\rangle$ by emitting a photon~\cite{Cohen-Tannoudjibook}.
 
 By further employing a rotating frame with the frequency of the laser field via the unitary operator
 $U(t)=\exp\left(-i/2\omega_{L}\sigma_{z}t\right)$, we obtain the time-independent Hamiltonian in Eq~\eqref{eq:evol}, which is derived from
 \begin{equation}
 	\label{eq:Hamitimdep3}
 	\mathcal{H}=i\hbar\frac{dU^{\dag}(t)}{dt}U(t)+U^{\dag}(t)\mathcal{H}_{\mathrm{RWA}}(t)U(t).
 \end{equation}
 In the rotating frame, the corresponding density matrix is defined as
  \begin{equation}
 	\label{eq:Hamitimdep3}
 	\rho(t)=U^{\dag}(t)\varrho(t)U(t),
 \end{equation}
and its dynamical evolution is governed by the quantum master equation in Eq.~\eqref{eq:evol}.

\section{Derivation of the time evolution of $n$th partial density matrix}
\label{sec:appApardenmat}

To study the photon emission of the driven two-level single-molecule system,  we rewrite the dynamical evolution for the density matrix in Eq.~\eqref{eq:evol}  as
\begin{equation}
	\label{eq:evolQJ}
	\frac{d}{dt}\rho(t)=\mathcal{C}\rho(t)+\mathcal{J}\rho(t),
\end{equation}
where the first term accounts for the coherent evolution
\begin{equation}
	\label{eq:cohevo}
	\mathcal{C}\rho(t)=-\frac{i}{\hbar}[\mathcal{H}_{\mathrm{eff}}\rho(t)-\rho(t)\mathcal{H}_{\mathrm{eff}}^{\dag}],
\end{equation}
with $\mathcal{H}_{\mathrm{eff}}=\mathcal{H}-(i/2)\hbar\Gamma\sigma_{+}\sigma_{-}$ being  the non-Hermitian effective Hamiltonian, and the second term is the quantum jump operator 
\begin{equation}
	\label{eq:disope}
	\mathcal{J}\rho(t)=\Gamma\sigma_{-}\rho(t)\sigma_{+},
\end{equation}
which is associated with the dissipation of the quantum system from the exited state $|e\rangle$ to the ground state $|g\rangle$.

In the interaction picture, the dynamical evolution of the density matrix
in Eq.~\eqref{eq:evolQJ} can be rewritten as
\begin{equation}
	\label{eq:evolQJintpic}
	\frac{d}{dt}\rho^{I}(t)=\mathcal{J}(t)\rho^{I}(t),
\end{equation}
where we have used the definitions
\begin{equation}
	\label{eq:intpicdef}
	\begin{split}
		\rho^{I}(t)&=\mathcal{T}^{\dag}(t,0)\rho(t),\\
		\mathcal{J}(t)&=\mathcal{T}^{\dag}(t,0)\mathcal{J}\mathcal{T}(t,0),
	\end{split}
\end{equation}
in terms of the time operator $	\mathcal{T}(t,0)=\exp\left(\mathcal{C}t\right)$.
Converting Eq.~\eqref{eq:evolQJintpic} into an integral form yields
\begin{equation}
	\label{equ12}
	\rho^{I}(t)=\rho(0)+\int_{0}^{t}dt'\mathcal{J}(t')\rho^{I}(t'),
\end{equation}
and subsequent iteration of this integral equation gives
\begin{equation}
	\label{equ13}
	\rho^{I}(t)=\rho(0)+\sum_{n=1}^{\infty}\int_{0}^{t}dt_{1}\cdot\cdot\cdot
	\int_{0}^{t_{n-1}}dt_{n}\mathcal{J}(t_{1})\cdot\cdot\cdot\mathcal{J}(t_{n})\rho(0).
\end{equation}

By transforming Eq.~(\ref{equ13}) back into the Schr\"{o}dinger picture, we can express the density matrix of the driven two-level system in Eq.~\eqref{eq:evolQJ} as
\begin{equation}
	\label{eq:redden1}
	\rho(t)=\sum_{n=0}^{\infty}\rho^{(n)}(t)=\rho^{(0)}(t)+\rho^{(1)}(t)+\cdots,
\end{equation}
where $\rho^{(n)}(t)$ are the $n$th partial density matrix related to the number of quantum jumps from the exited state $|e\rangle$ to the ground state $|g\rangle$ which describes no photon emission
\begin{equation}
	\label{eq:nojump}
	\rho^{(0)}(t)= \mathcal{T}(t,0)\rho(0),
\end{equation}
and the $n$ $(n>0)$ photons emission
\begin{equation}
	\label{eq:nthjump}
	\begin{split}
		\rho^{(n)}(t)&=\int_{0}^{t}dt_{1}\cdots\int_{0}^{t_{n-1}}dt_{n} \mathcal{T}(t,t_{1})\mathcal{J}\mathcal{T}(t_{1},t_{2})\cdots\\
		&\quad\times\mathcal{T}(t_{n-2},t_{n-1})\mathcal{J}\mathcal{T}(t_{n-1},t_{n})\rho(0),
	\end{split}
\end{equation}
respectively. Differentiating the $n$th partial density matrix in Eqs.~\eqref{eq:nojump} and~\eqref{eq:nthjump} with respect to time yields
\begin{equation}
	\label{eq:difrel1}
	\begin{split}
		\frac{d}{dt}\rho^{(0)}(t)&=\mathcal{C}\rho^{(0)}(t),\\
		\frac{d}{dt}\rho^{(n)}(t)&=\mathcal{C}\rho^{(n)}(t)+\mathcal{J}\rho^{(n-1)}(t).
	\end{split}
\end{equation}
By further setting $\rho^{(n)}(t)=0$ for $n<0$, we can thus obtain the time evolution of $n$th partial density matrix in Eq.~\eqref{eq:evopardenmat}.

\section{Solution for the $n$-th order derivatives of the generating function $\mathcal{P}(s,t)$ with respect to the counting variable $s$}
\label{sec:appBsolnthord}
To obtain the $n$-th order derivatives of the generating function $\mathcal{P}(s,t)$ with respect to the counting variable $s$, we can further construct a system of differential equations in matrix form as
\begin{equation}
	\label{appBeq:difequ}
	\begin{split}
		\frac{d}{dt}\mathscr{Y}(s,t)=\mathscr{M}(s)\mathscr{Y}(s,t),
	\end{split}
\end{equation}
where the column vector
\begin{equation}
	\label{appBeq:colvec}
	\mathscr{Y}(s,t)=\left
	(\begin{array}{c}
		\mathbb{Y}(s,t)\\
	   \frac{\partial}{\partial{s}}\mathbb{Y}(s,t)\\
		\vdots\\
		\frac{\partial^{n}}{\partial{s^{n}}}\mathbb{Y}(s,t)
	\end{array}
	\right),
\end{equation}
comprises 4$(n+1)$ components, including the generating functions $\mathcal{U}(s,t),\mathcal{V}(s,t),\mathcal{W}(s,t),\mathcal{P}(s,t)$ and their derivatives of orders from the 1st to the $n$-th with respect to the counting variable $s$, the initial condition satisfies $\mathscr{Y}(s,0)=\left(\mathbb{Y}(s,0),\cdots,(\partial^{k}/\partial{s^{k}})\mathbb{Y}(s,0),\cdots\right)^{\dag}$ with $(\partial^{k}/\partial{s^{k}})\mathbb{Y}(s,0)=0$, for $k=1,2,\cdots,n$, 
and the $4(n+1)\times4(n+1)$ coefficient matrix $\mathscr{M}(s)$ is expressed as
\begin{widetext}
	\begin{equation}
		\label{appBeq:cofmat}
		\mathscr{M}(s)=\left
		(\begin{array}{cccccccccc}
			\mathcal{M}(s)&0&0&\cdots&0&0&0\\
			\frac{\partial}{\partial s}\mathcal{M}(s)&\mathcal{M}(s)&0&\cdots&0&0&0\\
			\vdots&\vdots&\vdots&\ddots&\vdots&\vdots&\vdots\\
			0&0&0&\cdots&(n-1)\frac{\partial}{\partial s}\mathcal{M}(s)&\mathcal{M}(s)&0\\
			0&0&0&\cdots&0&n\frac{\partial}{\partial s}\mathcal{M}(s)&\mathcal{M}(s)		
		\end{array}
		\right),
	\end{equation}
\end{widetext}
with the $4\times4$ coefficient matrix $\mathcal{M}(s)$ in Eq.~\eqref{eq:coemat}.

The solution of the column vector $\mathscr{Y}(s,t)$ in Eq.~\eqref{appBeq:difequ} can be formally expressed as
\begin{equation}
	\label{appAeq:solcolvec}
		\mathscr{Y}(s,t)=\exp\left[\mathscr{M}(s)t\right]\mathscr{Y}(s,0).
\end{equation}
The $k$-th order derivative ($k=1,2,\cdots,n$)  of the generating function $\mathcal{P}(s,t)$ with respect to the counting variable $s$ corresponds to the $4(k+1)$-th component in the column vector $	\mathscr{Y}(s,t)$.
Consequently, several key statistical properties of photon emission, such as the probability the single-molecule system emits $n$ photons prior to time $t$ in Eq.~\eqref{eq:pronpho} and the $n$-th order factorial moment of the number of emitted photons in Eq.~\eqref{eq:kthmom}, can be derived from $(\partial^{n}/\partial{s^{n}})\mathcal{P}(s,t)$ by setting $s=0$ and $s=1$, respectively.

%

\end{document}